\documentclass{article}

\usepackage[a4paper, total={6.5in, 8.5in}]{geometry}

\usepackage{caption}
\usepackage{tikz}
\usetikzlibrary{arrows}
\usepackage[normalem]{ulem}
\usepackage{xcolor}
\usepackage{amssymb,amsmath,tabularx}

\newtheorem{theorem}{Theorem}
\newtheorem{lemma}{Lemma}

\newcommand{\U}{\mathit{U}}    
\newcommand{\SU}{\mathit{SU}}  

\newcommand{\bsv}{\boldsymbol{v}}

\newcommand{\bsx}{\boldsymbol{x}}

\newcommand{\bst}{\boldsymbol{t}}
\newcommand{\bsz}{\boldsymbol{z}}
\newcommand{\bsu}{\boldsymbol{u}}

\newcommand{\bsy}{\boldsymbol{y}}

\newcommand{\bsphi}{\boldsymbol{\phi}}
\newcommand{\bszero}{\boldsymbol{0}}
\newcommand{\rd}{\mathrm{d}}
\newcommand{\ri}{\mathrm{i}}

\newcommand{\calI}{{\mathcal{I}}}

\newcommand{\calO}{{\mathcal{O}}}
\newcommand{\calQ}{{\mathcal{Q}}}

\newcommand{\bbN}{{\mathbb{N}}}
\newcommand{\bbR}{{\mathbb{R}}}

\definecolor{darkred}{RGB}{139,0,0}
\definecolor{darkgreen}{RGB}{0,100,0}
\definecolor{darkmagenta}{RGB}{139,0,139}

\title{Lattice field computations via recursive numerical integration}

 \author{
     Tobias Hartung\footnote{%
       Computation-based Science and Technology Research Center, The Cyprus Institute, 20 Konstantinou Kavafi Street, 2121 Nicosia, Cyprus
       and
       Department of Mathematics, King's College London, Strand, London WC2R 2LS, United Kingdom
       (email: tobias.hartung@desy.de)},
     \quad
     Karl Jansen\footnote{%
       NIC, DESY Zeuthen, Platanenallee 6, 15738 Zeuthen, Germany
       (email: karl.jansen@desy.de)},
     \quad
     Frances Y.~Kuo\footnote{%
       School of Mathematics and Statistics, UNSW Sydney, Sydney NSW 2052, Australia
       (email: f.kuo@unsw.edu.au)},
     \\
     Hernan Le\"ovey\footnote{%
       Structured Energy Trading, AXPO Trading \& Sales, Parkstrasse 23, 5400 Baden, Germany
       (email: hernaneugenio.leoevey@axpo.com)},
     \quad
     Dirk Nuyens\footnote{%
       Department of Computer Science, KU Leuven, Celestijnenlaan 200A, 3001 Leuven, Belgium
       (email: dirk.nuyens@cs.kuleuven.be)},
     \quad
     Ian H.~Sloan\footnote{%
       School of Mathematics and Statistics, UNSW Sydney, Sydney NSW 2052, Australia
       (email: i.sloan@unsw.edu.au)}}

\date{December 2021 \\
Submitted to the Proceedings of the 38th International Symposium on Lattice Field Theory,
26th-30th July 2021, Zoom/Gather at Massachusetts Institute of Technology}

\begin{document}

\maketitle

\abstract{We investigate the application of efficient recursive numerical
integration strategies to models in lattice gauge theory from
quantum field theory. Given the coupling structure of the physics
problems and the group structure within lattice cubature rules for numerical
integration, we show how to approach these problems efficiently by means
of Fast Fourier Transform techniques. In particular, we consider
applications to the quantum mechanical rotor and compact $\U(1)$ lattice gauge
theory, where the physical dimensions are two and three. This proceedings article
reviews our results presented in J.\ Comput.\ Phys 443 (2021) 110527 \cite{HJKLNS21}. }

\section{Introduction} \label{sec:intro}

\emph{Lattice Field Theory} (LFT) serves as a non-perturbative tool to
regulate Feynman's path integral, by removing infinities in the infra-red
and ultra-volet. In LFT continuum models in physics are formulated on a
finite Euclidean space-time lattice with lattice spacing $h$. In this way,
an ultra-violet cut-off, $1/h$, is provided through a non-vanishing value
of the lattice spacing. At the same time, the finite volume equips LFTs
with an infra-red cut-off providing thus to a well-defined path integral,
see e.g., \cite{Gattringer:2010zz} for an introduction to lattice field
theories. (See \cite{Hartung:2018usn,Jansen:2019uei} for an alternative
approach using the $\zeta$-regularization.)

Another important aspect of Euclidean LFT is that the path integral and
physical observables can be computed numerically within this framework.
One very prominent example of a LFT is \emph{Lattice Quantum
Chromodynamics} (LQCD) which is the lattice version of \emph{Quantum
Chromodynamics} (QCD), the theory of the interaction between quarks and
gluons. LQCD has been extremely successful in computing e.g., the low
lying hadron spectrum, non-perturbative matrix elements and form factors,
fundamental parameters of QCD and non-zero temperature physics, see e.g.,
the \emph{Flavour Lattice Averaging Group} (FLAG) review
\cite{FlavourLatticeAveragingGroup:2019iem}.

The main tool for simulating LQCD are \emph{Markov Chain Monte Carlo}
(MCMC) methods such as Hybrid Monte Carlo, see \cite{Luscher:2010ae} for
an overview. The employed methods and algorithms have been immensely
improved over the last years, leading to large factors of speedup in the
numerical calculations. In addition, the structure of LQCD and of many
other LFTs is of local nature, coupling essentially only nearest
neighbours on the lattice. This allows for massively parallel simulations
on state of the art supercomputers with (hundred) thousands of processors,
which led to computations on lattices of size close to physical relevance,
meaning that the first two quark generations, i.e., the up, down, strange
and charm quarks, assume their physical values as determined from
experiments.

Despite this enormous success, the MCMC approach is unable to address
problems which lead to a complex integrand in the path integral, a
situation which is referred to as {\em sign problem} \cite{Troyer:2004ge}.
These problems are of fundamental nature in the context of high energy
physics. In particular, they are related to the questions: Why is there
more matter than anti-matter in the universe? Why is there more
experimentally detected CP violation than the standard model predicts?
What are the physics of topological systems?

In addition, MCMC methods suffer from the problem of auto-correlations
which is an inherent property of MCMC algorithms. This problem becomes
severe when the continuum limit in LQCD is taken, i.e., when the lattice
spacing is shrunk towards zero, see \cite{Albandea:2021lvl} and the
references therein. In particular, it has been noticed
\cite{Schaefer:2010hu} that the topological charge in LQCD starts to
freeze in Monte Carlo simulations, making it thus very hard to take the
continuum limit.

The above sketched shortcomings of MCMC methods has motivated researchers
working in LFTs to constantly look for alternative approaches. The very
large numerical demand of lattice simulations has triggered new approaches
to accelerate LFT calculations even more and to achieve a very high
accuracy in evaluating the path integral. The sign problem has led to new
developments in using numerical techniques such as tensor networks
\cite{Banuls:2019rao} or tensor renormalization group methods~
\cite{Akiyama:2021nhe}.

In this proceedings article we report on a particular attempt to address
the above mentioned problems by using efficient recursive integration
techniques, combined with a Fourier analysis, following our original work
in~\cite{HJKLNS21}. The idea is illustrated and applied to an abelian
$\U(1)$ compact lattice gauge theory in two dimensions. By applying
\emph{lattice cubature rules} (see e.g., \cite{DKS13,SJ94}), the
underlying group structure of this cubature allows for efficient
computation when plugged into integrands exhibiting special structure as
in the physics problems considered here. Although we will demonstrate the
method here in two dimensions only, we see the prospect to generalize our
method to higher dimensions and also to non-abelian groups in our future
work.

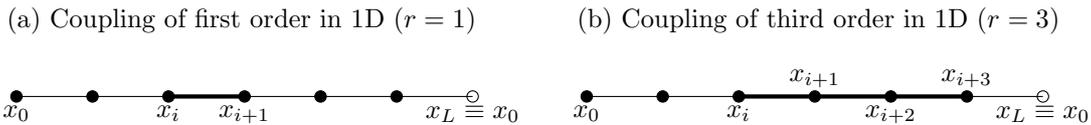
\begin{figure}
  \centering
  \begin{tikzpicture}[scale=1]
    \begin{scope}
      \node[anchor=west] at (-0.25,1) {(a) Coupling of first order in 1D $(r=1)$};
      \foreach \i in {0,...,5} {
        \filldraw[black] (\i,0) circle (.075);
      }
      \draw (6,0) circle (.075);
      \node[below] at (0,0) {$x_0$};
      \node[below] at (2,0) {$x_i$};
      \node[below] at (3,0) {$x_{i+1}$};
      \node[below] at (6,0) {$x_L \equiv x_0$};
      \draw (0,0) -- (6,0);
      \draw[ultra thick] (2,0) -- (3,0);
    \end{scope}
    \begin{scope}[shift={(7.5,0)}]
      \node[anchor=west] at (-0.25,1) {(b) Coupling of third order in 1D $(r=3)$};
      \foreach \i in {0,...,5} {
        \filldraw[black] (\i,0) circle (.075);
      }
      \draw (6,0) circle (.075);
      \node[below] at (0,0) {$x_0$};
      \node[below] at (2,0) {$x_i$};
      \node[above] at (3,0) {$x_{i+1}$};
      \node[below] at (4,0) {$x_{i+2}$};
      \node[above] at (5,0) {$x_{i+3}$};
      \node[below] at (6,0) {$x_L \equiv x_0$};
      \draw (0,0) -- (6,0);
      \draw[ultra thick] (2,0) -- (5,0);
    \end{scope}
  \end{tikzpicture}
\caption{\textbf{Different couplings in LQCD} \;(Figure taken from \cite[Fig.~1]{HJKLNS21})} \label{fig1}
\end{figure}

To be more concrete, we will initially consider $L$-dimensional integrals
(the ``dimensionality'' here refers to the number of integration variables
rather than the space/time dimensions) of the form
\begin{align} \label{eq:int1}
 \int_{D^L} \prod_{i=0}^{L-1} f_i\big(x_i,x_{i+1}\big) \,\rd\bsx,
 \qquad\mbox{with $\bsx = (x_0,\ldots,x_{L-1})$ and $x_L\equiv x_0$,}
\end{align}
where each variable $x_i$ belongs to a bounded domain $D\subset \bbR$ and
each function $f_i$ depends only on two consecutive variables $x_i$ and
$x_{i+1}$ -- which is called a {\em first order coupling} --
and we implicitly assume periodic boundary conditions.
\ref{fig1} sketches first order as well as third order coupling.

As already said above, we propose to solve lattice systems with such a
coupling structure by approximating  the involved integrals employing
recursive numerical integration methods, where the lattice cubature rules
used belong to a special family among \emph{Quasi-Monte Carlo} rules (see
e.g., \cite{DKS13,DP10,Hic98b,Lem09,Nie92,Nuy14,SJ94}). The work we
describe here is part of an ongoing effort by some of us to apply
alternative to MCMC mathematical methods to tackle lattice models;
previous works employ the Quasi-Monte Carlo approach \cite{JLAGM14},
polynomially exact integration rules \cite{GHJLV16a,Hartung:2020uuj}, and
a first applications of the recursive numerical integration technique to
address quantum mechanical models \cite{AGHJLV16,HJKLNS21}. Quasi-Monte
Carlo methods have also been considered in
\cite{Borowka:2018goh,deDoncker:2018nqe} for multiloop calculations in
perturbation theory.

In the next section we will provide the physical models considered. In the
subsequent sections we will outline the efficient methods and the
corresponding computational complexities. Executable Julia codes and
numerical results can be found in the original work \cite{HJKLNS21}.

\section{Description of physical models} \label{sec:physics}

In this section we will introduce the quantum mechanical rotor
\cite{AGHJLV16,Bietenholz:1997kr,Bietenholz:2010xg} and a $2$-dimensional
compact $\U(1)$ lattice gauge theory, see e.g., \cite{Gattringer:2010zz}
for an introduction into lattice field theories.

\subsection{Quantum rotor} \label{sec:rotor}

The quantum rotor is a quantum mechanical model which describes a particle
with mass $m_0$ moving on a circle with radius~$r_0$, see e.g.,
\cite{AGHJLV16,Bietenholz:1997kr,Bietenholz:2010xg}. We can supply the
particle with a moment of inertia $I = m_0 r_0^2$. The quantum rotor can
be formulated on a time lattice with lattice spacing $h = T/L$ between
neighbouring lattice points, where $T$ is the final time and $L$ is the
number of lattice points. The variables of the system are taken to be
angles $\phi_i\in D=[-\pi,\pi)$ defined at each time lattice point. In the
continuum, the action of the system is given by
\begin{align} \label{eq:limit}
 S(\phi)=\int_{0}^T \frac{I }{2}\Big(\frac{\rd \phi}{\rd t}\Big)^2 \rd t.
\end{align}
The discretized version of the action of the quantum rotor, using $\frac{1}{2}(\frac{\rd \phi}{\rd
t})^2 \approx \frac{1}{2}(\frac{\phi_{i+1} - \phi_i}{h})^2 \approx \frac{1
- \cos(\phi_{i+1} - \phi_i)}{h^2}$, is then given by
\begin{align*} 
  S[\bsphi]
  \,=\, \frac{I}{h^2}\sum_{i=0}^{L-1} \big(1-\cos(\phi_{i+1}-\phi_{i})\big)
\end{align*}
and we impose periodic boundary conditions. The choice of the action will
allow us to employ a combination of techniques as detailed below to make
use of the Fast Fourier transform (FFT), reducing significantly the
computational cost of the problem. As an observable we will take
\begin{align*}
  O[\bsphi] \,=\, \cos(\phi_{k+1} - \phi_k) \quad\mbox{for any $k$},
\end{align*}
which is evaluated through the path integral as
\begin{align} \label{eq:rotor}
  \langle O[\bsphi] \rangle \,=\, \frac
  {\int_{D^L} \cos\big(\phi_{k+1} - \phi_k) \, \exp(\beta \sum_{i=0}^{L-1}\cos(\phi_{i+1}-\phi_{i})\big) \,\rd\bsphi}
  {\int_{D^L} \exp\big(\beta\sum_{i=0}^{L-1}\cos(\phi_{i+1}-\phi_{i})\big) \,\rd\bsphi},
\end{align}
where $\beta= \frac{I}{h^2}= \frac{I L^2}{T^2}$. In order to evaluate the
above integrals we will in the following convert their domain into the
unit cube $[0,1)^L$ since the integration domain $[0,1)^L$ is the
classical domain where Quasi-Monte Carlo rules are defined.

\subsection{Quantum compact abelian gauge theory} \label{sec:QED}

Gauge theories are at the heart of the standard model of high energy
physics. They are constructed such that they exhibit local gauge
invariance, i.e., physical observables and the action of gauge models are
invariant under local changes of the field variables defining a gauge
theory. Using gauge invariant theories in constructing the standard model
has been extremely successful. In particular, with the discovery of the
Higgs boson, all particles predicted by the standard model have been
identified experimentally. Thus, we now have a complete microscopic
description of the interaction of the fundamental particles building all
matter.

There are a number of non-perturbative aspects of gauge theories, such as
confinement, topological effects and the existence of glueballs. The main
tool, as formulated by Wilson \cite{Wilson:1974sk} to understand these
non-perturbative phenomena is {\em Lattice Gauge Theory} (LGT). On the
lattice, the fundamental {\em gauge fields} are group valued and mainly
taken from the abelian group $\U(1)$, or the non-abelian ones, $\U(N)$ or
$\SU(N)$ with $N\ge 2$.

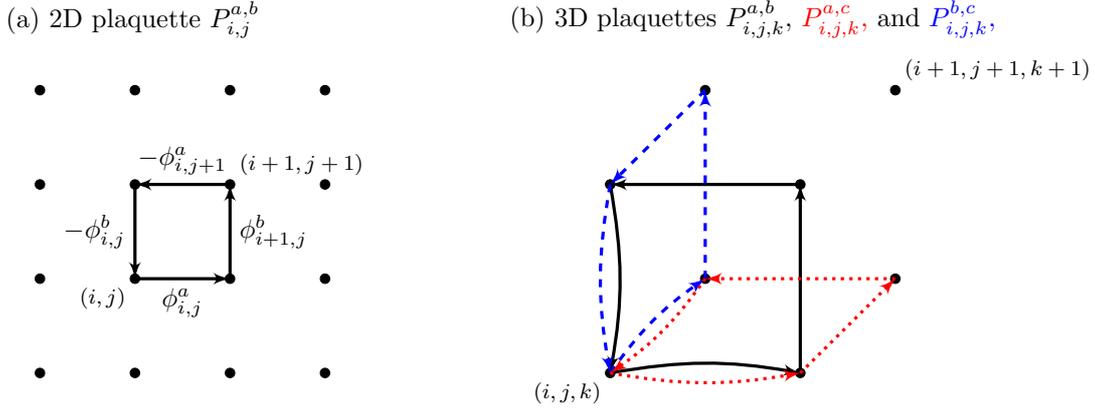
\begin{figure}
  \centering
  \begin{tikzpicture}[scale=1.25,>=latex']
    \node at (-.025,3.75) {(a) 2D plaquette $P_{i,j}^{a,b}$};
    \foreach \i in {-1,...,2}{
      \foreach \j in {0,...,3}{
        \filldraw[black] (\i,\j) circle (.05);
      }
    }
    \draw[->,black,very thick] (0,1) -- (1,1);
    \draw[->,black,very thick] (1,1) -- (1,2);
    \draw[->,black,very thick] (1,2) -- (0,2);
    \draw[->,black,very thick] (0,2) -- (0,1);
    \node[black,below] at (0.5,1) {$\phi_{i,j}^a$};
    \node[black,above] at (0.5,2) {$-\phi_{i,j+1}^a$};
    \node[black,left] at (0,1.5) {$-\phi_{i,j}^b$};
    \node[black,right] at (1,1.5) {$\phi_{i+1,j}^b$};
    \node[black,below left] at (0,1) {\footnotesize $(i,j)$};
    \node[black,above right] at (1,2) {\footnotesize $(i+1,j+1)$};
    \node at (6.5,3.75) {(b) 3D plaquettes {\color{black} $P_{i,j,k}^{a,b}$, {\color{red} $P_{i,j,k}^{a,c}$,} and {\color{blue} $P_{i,j,k}^{b,c}$,}}};
    \filldraw[black] (5,0) circle (.05);
    \filldraw[black] (7,0) circle (.05);
    \filldraw[black] (5,2) circle (.05);
    \filldraw[black] (7,2) circle (.05);
    \filldraw[black] (6,1) circle (.05);
    \filldraw[black] (8,1) circle (.05);
    \filldraw[black] (6,3) circle (.05);
    \filldraw[black] (8,3) circle (.05);
    \node[black,below left] at (5,0) {\footnotesize $(i,j,k)$};
    \node[black,above right] at (8,3) {\footnotesize $(i+1,j+1,k+1)$};
    \draw[->,black,very thick] (5,0) to[out=10,in=170] (7,0);
    \draw[->,black,very thick] (7,0) -- (7,2);
    \draw[->,black,very thick] (7,2) -- (5,2);
    \draw[->,black,very thick] (5,2) to[out=280,in=80] (5,0);
    \draw[->,dashed,blue,very thick] (5,0) to[out=55,in=215] (6,1);
    \draw[->,dashed,blue,very thick] (6,1) -- (6,3);
    \draw[->,dashed,blue,very thick] (6,3) -- (5,2);
    \draw[->,dashed,blue,very thick] (5,2) to[out=260,in=100] (5,0);
    \draw[->,dotted,red,very thick] (5,0) to[out=-10,in=190] (7,0);
    \draw[->,dotted,red,very thick] (7,0) -- (8,1);
    \draw[->,dotted,red,very thick] (8,1) -- (6,1);
    \draw[->,dotted,red,very thick] (6,1) to[out=235,in=35] (5,0);
  \end{tikzpicture}
\caption{\textbf{How to construct a plaquette in two 
and three 
dimensions} \\ (Figure taken from \cite[Fig.~2]{HJKLNS21})
In the two dimensional case, we construct the plaquette $P_{i,j}^{a,b}$ at
lattice point $(i,j)$ by carrying the field along the
path $(i,j)\to(i+1,j)\to(i+1,j+1)\to(i,j+1)\to(i,j)$.
The contribution of the angles $\phi$ is of positive sign if moving right or up, and
the contribution is of negative sign if moving left or down. We index left/right movements
with a superscript~$a$ and up/down movements with superscript $b$.
For the three dimensional case, we add an extra forward/backward movement, which is denoted by
the superscript $c$. This forms three independent smallest
loops that can be taken:  the $(a,b)$-plane (black solid line),
the $(a,c)$-plane (red dotted line), and the $(b,c)$-plane (blue dashed line).
One can see that each of these planes contribute equally with a ``2-dimensional'' plaquette.
All visualized links are straight lines, they are only bent for
better visualization. }\label{fig:QED-plaquette}
\end{figure}

Ideally, a study of gauge theories in three space and one time dimension
($3+1$ dimensions) would be preferable. However, this is presently not
affordable within our approach and we therefore resort to a lower
($1+1$)-dimensional model, the ($1+1$)-dimensional compact $\U(1)$ lattice
gauge theory. The path integral of this model reads
\begin{align} \label{eq:QED2D}
  \int_{D^{2L^2}}
  \exp\bigg(\beta \sum_{i=0}^{L-1} \sum_{j=0}^{L-1}
  \cos\Big(\phi^a_{i,j} + \phi^b_{i+1,j} - \phi^a_{i,j+1} - \phi^b_{i,j} \Big)
  \bigg)
  \,\rd\bsphi.
\end{align}
Here $D=[-\pi,\pi]$, and we have parametric periodicity where all indices
should be taken modulo $L$. A typical observable to be considered is the
{\em plaquette expectation value}, see \ref{fig:QED-plaquette}(a),
\begin{align*}
  \cos\Big(\phi^a_{0,0} + \phi^b_{1,0} - \phi^a_{0,1} - \phi^b_{0,0} \Big),
\end{align*}
which we also study here.

The model above uses the plaquette -- the smallest Wilson loop -- to
define the action. Alternative forms of the action, employing larger
Wilson loops can be considered and are actually used in lattice
simulations as they can improve the extrapolation behavior to the
continuum limit, i.e., the limit where the lattice spacing is sent to
zero. The only requirement is that these Wilson loops are gauge invariant
and that the corresponding lattice action converges to the continuum
action when the lattice spacing is sent to zero. Our method generalizes
straightforwardly to such extended Wilson loops and in our original work
\cite{HJKLNS21} there is a detailed discussion of such a situation.

The construction of a $\U(1)$ compact lattice gauge theory can be
generalized to three dimensions as illustrated in
\ref{fig:QED-plaquette}(b). The path integral then reads
\begin{align*} 
  &\int_{D^{3L^3}}
  \exp\bigg(\beta \sum_{i=0}^{L-1} \sum_{j=0}^{L-1} \sum_{k=0}^{L-1} \bigg[
  \cos\Big(\phi^a_{i,j,k} - \phi^a_{i,j+1,k} - \phi^b_{i,j,k} + \phi^b_{i+1,j,k} \Big) \nonumber\\
  &\qquad\qquad\qquad\qquad\qquad\qquad
  + \cos\Big(\phi^c_{i,j,k} - \phi^c_{i+1,j,k} - \phi^a_{i,j,k} + \phi^a_{i,j,k+1} \Big) \nonumber\\
  &\qquad\qquad\qquad\qquad\qquad\qquad
  + \cos\Big(\phi^b_{i,j,k} - \phi^b_{i,j,k+1} - \phi^c_{i,j,k} + \phi^c_{i,j+1,k} \Big)
  \bigg] \bigg)
  \,\rd\bsphi,
\end{align*}
while for the observable we again take the plaquette expectation value
\[
  O[\bsphi]
  \,=\, \cos\Big(\phi^a_{0,0,0} - \phi^a_{0,1,0} - \phi^b_{0,0,0} + \phi^b_{1,0,0}\Big).
\]

In a more general way, the plaquette variable at a point $(i,j)$ can be
written as
\begin{align}\label{eq:plaquette-definition}
 P_{i,j}^{a,b} = \Phi_{i,j}^a\Phi_{i+1,j}^b(\Phi_{i,j+1}^a)^{-1}(\Phi_{i,j}^b)^{-1}
 \equiv e^{\ri\left(\phi_{i,j}^a+\phi_{i+1,j}^b-\phi_{i,j+1}^a-\phi_{i,j}^b\right)}.
\end{align}
This generalized form allows us to formulate the path integral and the
gauge invariant observables by taking $\Phi_{i,j,\ldots}^\alpha$ from
$\U(N)$ or $\SU(N)$ and taking the trace as well as real parts of
\eqref{eq:plaquette-definition}. For the physically relevant cases of the
standard model, namely the weak interaction, or QCD (quantum
chromodynamics or strong nuclear force), the relevant groups are $\SU(2)$
and $\SU(3)$, respectively.

\section{Recursive numerical integration and complexity results} \label{sec:recur}

We present in this section a strategy to approximate the integral
\eqref{eq:int1} using iterated integration and then replacing each of the
iterated integrals by a numerical quadrature/cubature. We refer to this
method as \textit{recursive numerical integration}. The method first uses
the fact that the integral \eqref{eq:int1} can be rewritten as
\begin{align} \label{eq:int2}
    \calI
    &\,=\, \int_D\cdots\int_D
    f_0\big(x_0,x_1\big)
    f_1\big(x_1,x_2\big)f_2\big(x_2,x_3\big)
    \cdots
    f_{L-1}\big(x_{L-1},x_0\big) \,\rd x_0 \cdots \rd x_{L-1} \nonumber\\
    &\,=\, \int_D \bigg[\int_D\cdots
    \left(\int_D
    \left(\int_D f_0\big(x_0,x_1\big) f_1\big(x_1,x_2\big)\,\rd x_1\right)
    f_2\big(x_2,x_3\big)
    \,\rd x_2 \right)
    \cdots f_{L-1}\big(x_{L-1},x_0\big) \,\rd x_{L-1} \bigg] \,\rd x_0,
\end{align}
with  $\bsx = (x_0,\ldots,x_{L-1})$ and $x_L\equiv x_0$. Due to the
periodicity assumption of the indices of the variables, i.e, $x_L \equiv
x_0$, we can start the iterated integration by choosing any variable of
preference. The integrand presented here exhibits what is called a
\textit{first order coupling}, which means that each variable $x_i$ is
coupled via the physical model to its nearest neighboring variables
$x_{i-1}$ and $x_{i+1}$ in a way that only these three variables are
present in the iterated integration of the variable $x_i$.

Recursive integration has been considered in e.g.,
\cite{Cra08,GenKal86,Hay06,Hay11}, but not for integrands with parametric
periodicity $x_L \equiv x_0$.

If we have an observable function that (i) depends only on one variable,
or (ii) depends on two consecutive variables as in the factors $f_i$, or
(iii) depends on many variables but takes the form of a product of first
order couplings as for the integrand in \eqref{eq:int1}, we will also
arrive at an integral that can be expressed iteratively in the same manner
as in \eqref{eq:int2}. For all physical models considered in this
proceedings article, we have observable functions that affect at most two
consecutive variables (or can be decomposed into a direct sum of such
functions) and therefore we can condense the effect of the observable
function into a new integrand factor which, without loss of generality,
can be included with the index $0$ and be written as $f_0^\star
\big(x_0,x_1\big):= O[x_0,x_1]f_0\big(x_0,x_1\big)$. In the coming
sections, when the observable function is included in the integration, we
will skip the $^\star$~notation and just mention that the integrand $f_0$
may be different from the other integrands $f_i$ involved.

When the iterated integrals are replaced by the same approximating
quadrature rule with points $t_0,\ldots,t_{n-1}\in D$ and weights
$w_0,\ldots,w_{n-1}\in \bbR$, the resulting approximation scheme takes the
form
\begin{align} \label{eq:prod2}
    \calQ
    &\,=\, \sum_{k_0=0}^{n-1} w_{k_0} \bigg[
    \sum_{k_{L-1}=0}^{n-1} w_{k_{L-1}} \cdots \bigg(\sum_{k_2=0}^{n-1} w_{k_2}
    \bigg(\sum_{k_1=0}^{n-1} w_{k_1} f_0\big(t_{k_0},t_{k_1}\big) f_1\big(t_{k_1},t_{k_2}\big)\bigg)
    f_2\big(t_{k_2},t_{k_3}\big)\bigg) \nonumber\\
    &\hspace{10cm}
    \cdots
    f_{L-1}\big(t_{k_{L-1}},t_{k_0}\big) \bigg].
\end{align}

The recursive approximation procedure is then carried out as follows.
Define $M_i$ to be the $n\times n$ matrix with entries
\begin{align} \label{eq:Mi}
  (M_i)_{p,q} \,=\, f_i(t_p,t_q) \qquad\mbox{for}\quad p,q=0,\ldots,{n-1},
\end{align}
and let $W$ denote the $n\times n$ diagonal matrix with the weights
$w_0,\ldots,w_{n-1}$ on the diagonal. Now we can express the innermost
sum in \eqref{eq:prod2} as
\begin{align*}
  \sum_{k_1=0}^{n-1} w_{k_1} f_0\big(t_{k_0},t_{k_1}\big) f_1\big(t_{k_1},t_{k_2}\big)
  &\,=\, \sum_{k_1=0}^{n-1} (M_0)_{k_0,k_1}w_{k_1}^{1/2} \, w_{k_1}^{1/2} (M_1)_{k_1,k_2} \\
  &\,=\, \sum_{k_1=0}^{n-1} (M_0 W^{1/2})_{k_0,k_1}\, (W^{1/2} M_1)_{k_1,k_2}
  \,=\, (M_0 W M_1)_{k_0,k_2}.
\end{align*}
Further we have
\begin{align*}
 \sum_{k_2=0}^{n-1} w_{k_2}
 \bigg(\sum_{k_1=0}^{n-1} w_{k_1} f_0\big(t_{k_0},t_{k_1}\big) f_1\big(t_{k_1},t_{k_2}\big)\bigg)
 f_2\big(t_{k_2},t_{k_3}\big)
  &\,=\,
  \sum_{k_2=0}^{n-1} w_{k_2} (M_0 W M_1)_{k_0,k_2} (M_2)_{k_2,k_3} \\
  &\,=\, (M_0 W M_1 W M_2)_{k_0,k_3}.
\end{align*}
This leads to
\begin{align} \label{eq:fast1}
 \calQ
 &\,=\, \sum_{k_0=0}^{n-1} w_{k_0} (M_0 W M_1 W M_2 W \cdots M_{L-1})_{k_0,k_0} \nonumber\\
 &\,=\, \sum_{k_0=0}^{n-1} (W^{1/2} M_0 W M_1 W M_2 W \cdots M_{L-1} W^{1/2})_{k_0,k_0} \nonumber\\
 &\,=\, {\rm trace}(B),
 \qquad\mbox{with}\qquad B \,=\, W^{1/2} M_0 W M_1 W M_2 W \cdots M_{L-1} W^{1/2}.
\end{align}
The equation \eqref{eq:fast1} defines what we call the \textit{recursive
numerical integration} approach. It shows a clear advantage over the naive
approach that could be used for approximation in \eqref{eq:prod2} since
the naive approach can be interpreted as a product rule over the $L$
dimensional domain. If the quadrature rules has $n$ points, this implies
that the naive approach would have a cost equal to $n^L$, i.e.,
exponential in the dimension $L$. On the contrary, applying the recursive
numerical integration approach we have an evaluation procedure for the
approximation of the integral that reduces the complexity significantly
and takes advantage of the particular structure of the matrices $M_i$. The
matrix structure is of course a consequence of the structure of the
underlying integrand factors $f_i$ and the chosen quadrature rule.

\ref{tab1} reviews the cases considered in the work \cite{HJKLNS21}
for different possibilities of matrices $M_i$ one can find in LQCD
applications.

There are particular conditions on the integrand and quadrature rule that
lead to great improvements in terms of computational cost. In particular,
if
\begin{enumerate}
\item each function $f_i$ depends only on the difference of the two
    arguments, i.e., $f_i(u,v) = \kappa_i(v-u)$ for some function
    $\kappa_i:D\to\bbR$, and
\item each function $\kappa_i$ is \emph{periodic}, and
\item we have equally spaced points with equal weights $1/n$ (i.e., we
    have the rectangle rule),
\end{enumerate}
then the matrix $M_i$ is \emph{circulant} and therefore FFT can be used to
obtain the eigenvalues, see Scenarios~(A5)--(A7) in \ref{tab1}.

If we replace the domain $D$ in \eqref{eq:int1} by an $s$-dimensional
domain $D^s$,
\begin{align} \label{eq:intLs}
 \calI
 &\,=\, \int_{D^s} \cdots \int_{D^s}
 \prod_{i=0}^{L-1} f_i\big(\bsx_i,\bsx_{i+1}\big) \,\rd\bsx_0\cdots\rd\bsx_{L-1},
\end{align}
where $\bsx_i = (x_{i,0},\ldots,x_{i,s-1})\in D^s$, with
\[
  x_{i,j} \equiv x_{i\bmod L,\; j\bmod s} \qquad \mbox{for all}\quad i,j\in\bbN,
\]
then the one-dimensional quadrature rule in \eqref{eq:prod2} becomes an
$s$-dimensional cubature rule with points $\bst_0,\ldots,\bst_{n-1} \in
D^s$ and weights $w_0,\ldots,w_{n-1}\in\bbR$, and the matrices $M_i$ in
\eqref{eq:Mi} become
\begin{align*}
  (M_i)_{p,q} \,=\, f_i(\bst_p,\bst_q) \qquad\mbox{for}\quad p,q=0,\ldots,{n-1}.
\end{align*}
Scenarios~(A0)--(A4) from \ref{tab1} apply again in this case. The
following conditions are sufficient to ensure circulant matrices for the
more favorable Scenarios~(A5)--(A7):
\begin{enumerate}
\item each function $f_i$ depends only on the difference of the two
    arguments, i.e., $f_i(\bsu,\bsv) = \kappa_i(\bsv-\bsu)$ for some
    function $\kappa_i : D^s \to \bbR$, and
\item each function $\kappa_i$ is \emph{periodic with respect to each
    of the $s$ components}, and
\item we have a \emph{lattice cubature rule} with points
\[
 \bst_k = \frac{k\bsz\bmod n}{n}\qquad\mbox{for}\quad k=0,\ldots,n-1,
\]
and equal weights $1/n$.
\end{enumerate}

A lattice cubature rule has an additive group structure. This means that
the difference of two lattice points is another lattice point. Since the
lattice cubature rule has equal weights $1/n$, we obtain circulant
matrices $M_i$. This particular fact is the main motivation for favoring
lattice cubature rules above all other cubature rules. The cost in all
scenarios is \emph{independent of $s$}. Nevertheless, the error is
$\calO(n^{-\alpha})$, where $\alpha$ is determined by the cubature rule
and the implied constant may depend on~$s$.

\begin{table}[t]
\captionsetup{singlelinecheck=off} \caption[foo bar]{\textbf{Recursive
numerical integration for first order couplings} \;(Table taken from
\cite[Table~1]{HJKLNS21})
\begin{itemize}
 \item $M_i$ is the $n\times n$ matrix of $f_i$ at quadrature points.
 \item  $W$ is an $n\times n$ diagonal matrix with quadrature weights
     on the diagonal.
 \item {\tt eig} returns a diagonal matrix of eigenvalues.
 \item {\tt fft} takes the first column of a circulant matrix and
     returns a diagonal matrix of eigenvalues.
 \item The quadrature error for all cases is $\calO(n^{-\alpha})$,
     with $\alpha$ given by the performance of the quadrature rule.
 \item The scenarios and strategies extend to an $L$-fold product of
     $s$-dimensional integrals with the quadrature rule replaced by an
     $s$-dimensional cubature rule. The cost remains independent
     of~$s$. The error is again $\calO(n^{-\alpha})$, with $\alpha$
     given by the performance of the cubature rule, and with an
     implied constant dependent on $s$.
\end{itemize}
}
 \label{tab1}
 \begin{center}
 \begin{tabular}{|l|l|c|}
 \hline
 Scenario & Strategy & Cost \\
 \hline
 \hline
 (A0) naive implementation &
 $\begin{array}{l}
 \calQ = \mbox{direct product calculation}
 \end{array}$
 & $n^L$ \\
 \hline
 (A1) recursive integration &
 $\begin{array}{l}
  B = W^{1/2} M_0 W M_1 W \cdots M_{L-1} W^{1/2} \\
  \calQ = \sum_{k=0}^{n-1} B_{k,k}
 \end{array}$
 & $L\,n^{3}$ \\
 & & \vspace{-0.4cm} \\
 \hline
 (A2) $M_i = M$ &
 $\begin{array}{l}
  A = W^{1/2} M W^{1/2} \\
  B = A^L \\
  \calQ = \sum_{k=0}^{n-1} B_{k,k}
 \end{array}
 $
 & $\log(L)\,n^{3}$ \\
 & & \vspace{-0.4cm} \\
 \hline
 (A3) $M_i=M$ diagonalizable &
 $\begin{array}{l}
  A = W^{1/2} M W^{1/2} \\
  \Lambda = {\tt eig}(A) \\
  \calQ = \sum_{k=0}^{n-1} \Lambda_{k,k}^L
 \end{array}
 $
 & $n^{3}$ \\
 & & \vspace{-0.4cm} \\
 \hline
 (A4) $M_i = M$ except $M_0$ &
 $\begin{array}{l}
  A = W^{1/2} M W^{1/2} \\
  B = W^{1/2} M_0 W^{1/2} A^{L-1} \\
  \calQ = \sum_{k=0}^{n-1} B_{k,k}
 \end{array}
 $
 & $\log(L)\,n^{3}$ \\
 & & \vspace{-0.4cm} \\
 \hline
 (A5) $M_i$ circulant &
 $\begin{array}{l}
  \Lambda_i = {\tt fft}(M_i/n) \mbox{ for each $i$} \\
  \calQ = \sum_{k=0}^{n-1} \prod_{i=0}^{L-1} (\Lambda_i)_{k,k}
 \end{array}
 $
 & $L\,n\log(n)$ \\
 & & \vspace{-0.4cm} \\
 \hline
 (A6) $M_i = M$ circulant &
 $\begin{array}{l}
  \Lambda = {\tt fft}(M/n) \\
  \calQ = \sum_{k=0}^{n-1} \Lambda_{k,k}^L
 \end{array}
 $
 & $n\log(n)$ \\
 & & \vspace{-0.4cm} \\
 \hline
 \begin{tabular}{l}
 \!\!\!(A7) $M_i = M$ except $M_0$ \\
 \quad\;\, all circulant
 \end{tabular}
 &
 $\begin{array}{l}
  \Lambda_0 = {\tt fft}(M_0/n) \\
  \Lambda = {\tt fft}(M/n) \\
  \calQ = \sum_{k=0}^{n-1} (\Lambda_0)_{k,k}\, \Lambda_{k,k}^{L-1}
 \end{array}
 $
 & $n\log(n)$ \\
 \hline
\end{tabular}
\end{center}
\end{table}

\clearpage 

These strategies have been extended in \cite{HJKLNS21} to \textit{higher
order couplings} of order $r$, of the form
\begin{align*} 
 \calI
 &\,=\, \int_{D^L} \prod_{i=0}^{L-1} f_i\big(x_i,x_{i+1},\ldots,x_{i+r}\big)
 \,\rd \bsx \\
 &\,=\, \int_D\cdots\int_D
 f_0\big(x_0,x_1,\ldots,x_r\big)f_1\big(x_1,x_2,\ldots,x_{r+1}\big)
 \cdots f_{r}\big(x_{r},x_{r+1},\ldots,x_{2r}\big) \nonumber\\
 &\qquad\qquad\qquad\qquad\qquad\qquad\qquad\qquad\qquad\qquad
 \cdots
 f_{L-1}\big(x_{L-1},x_0,x_1,\ldots,x_{r-1}\big) \,\rd x_0 \cdots \rd x_{L-1}. \nonumber
\end{align*}
The trick is to group successive $r$ functions into a new factor to form a
product of $L/r$ factors (assuming for simplicity here that $L$ is a
multiple of $r$), and then apply the recursive strategies to these new
factors. This yields analogous results to Scenarios~(A1)--(A7), which are
denoted by Scenarios (B1)--(B7) in \cite[Table~2]{HJKLNS21}. In all
scenarios the error is again $\calO(n^{-\alpha})$, with $\alpha$
determined by the quadrature/cubature rule, and the implied constant now
depends on $r$.

\section{Applications} \label{sec:app-rotor}

In this section we consider applications of recursive numerical integration to the
quantum rotor and the 2D compact $\U(1)$ LGT problem. The interested reader can find numerical
experiments for these two models in our recent work \cite{HJKLNS21}.

\subsection{The Quantum rotor}

For the quantum rotor, the integrals for the numerator and denominator
of the normalized path integral ratio
\eqref{eq:rotor} are of the form
\[
  \int_{D^L} \prod_{i=0}^{L-1} f_i(x_{i+1}-x_i)\,\rd\bsx,
\]
where we have, after a change of variables,  $D = [0,1]$ and
\[
  f_i(x) = f(x) = \exp(\beta\cos(2\pi x))
  \qquad\mbox{for all } i=0,\ldots, L-1.
\]
The only exception is that in the numerator integral we replace $f_0$ by
\[
  f_0(x) = \cos(2\pi x)\, \exp(\beta\cos(2\pi x)).
\]

With a slight abuse of notation comparing with \eqref{eq:int2}, here we
have the (special) case $f_i(u,v) = \kappa_i(v-u) \equiv f_i(v-u)$, which
means that each integrand factor $f_i(u,v)$ can be considered in fact as a
function of one single variable by taking the difference of the two
arguments $u,v$. The resulting functions $\kappa_i \equiv f_i$ are
periodic and therefore we know that for the rectangle quadrature rule we
have Scenario~(A7) in \ref{tab1}, and thus the numerical integration
cost becomes $\calO(n\log(n))$.

The interesting case of higher order couplings for the quantum rotor arise
when we consider higher order finite difference approximations of the
physical discretization of the action in continuum~\eqref{eq:limit}. For
example, the central difference formula $(- x_{i+2} + 8x_{i+1}-8x_{i-1}+
x_{i-2})/(12h)$ of order~$h^4$ leads to (now with $\beta = IL^2/(144T^2)$)
\[
  \int_{D^L} \prod_{i=0}^{L-1} f\big(-x_{i+2} + 8x_{i+1}-8x_{i-1}+ x_{i-2}\big)\,\rd\bsx,
\]
which has order $r=4$. This case can be solved using $4$-dimensional
lattice cubature rules combined with recursive numerical integration,
following \cite[Scenario~(B4) in Table~2]{HJKLNS21}.

\subsection{The 2D compact $\U(1)$ lattice gauge theory model}

Now we consider the model \eqref{eq:QED2D} which can be expressed in the
generic form
\begin{align*}
  \calI
  &\,=\,
  \int_{D^{L^2}} \int_{D^{L^2}} \prod_{i=0}^{L-1} \prod_{j=0}^{L-1}
  f_{i,j}\left(x^a_{i,j} - x^a_{i,j+1} - x^b_{i,j} + x^b_{i+1,j} \right)
  \,\rd\bsx^a \,\rd\bsx^b.
\end{align*}
After a change of variables, we have $D = [0,1]$ and
\[
  f_{i,j}(x) = f(x) = \exp(\beta\cos(2\pi x))
  \qquad\mbox{for all } i,j=0,\ldots, L-1.
\]
To include the observable, in the numerator integral we will replace
$f_{0,0}$ by
\[
  f_{0,0}(x) = \cos(2\pi x)\, \exp(\beta\cos(2\pi x)).
\]
All these functions are clearly periodic. Below we will outline three
stages to simplify the problem.

\subsubsection*{STAGE ONE}

By rearranging the variables in the $a$-direction and the $b$-direction,
we can write
\begin{align*}
  \calI
  &\,=\,
  \int_{D^{L^2}} \prod_{i=0}^{L-1}
  \bigg(
  \underbrace{
  \int_{D^L} \prod_{j=0}^{L-1}
  f_{i,j}\left(x^a_{i,j} - x^a_{i,j+1} - x^b_{i,j} + x^b_{i+1,j} \right)
  \,\rd\bsx^a_i}_{\mbox{$=:\,g_i\left(\bsx^b_{i+1}-\bsx^b_i\right)$}}
  \bigg)
  \,\rd\bsx^b,
\end{align*}
where we used the fact that each factor over the index $i$ depends only on
$\bsx^a_i$, and wrote the integral over $\bsx^a\in D^{L^2}$ as a product
of integrals over $\bsx^a_i = (x^a_{i,0},\ldots,x^a_{i,L-1})\in D^L$. This
leads to
\begin{align} \label{eq:outer}
  \calI
   &\,=\,
  \int_{D^L} \cdots \int_{D^L}\prod_{i=0}^{L-1}
  g_i\left(\bsy_{i+1}-\bsy_i\right)
  \,\rd\bsy_0 \cdots \rd\bsy_{L-1},
\end{align}
where
\begin{align} \label{eq:inner}
  g_i(\bsy)
  &\,:=\,
  \int_{D^L} \prod_{j=0}^{L-1}
  f_{i,j}\left(x_j - x_{j+1} + y_j\right)
  \,\rd\bsx.
\end{align}

Thus we have obtained a nested integration problem where for the outer
integral \eqref{eq:outer} we have an integrand with first order couplings
of the form \eqref{eq:intLs} (with $s$ replaced by $L$), and for the inner
integral \eqref{eq:inner} we have first order couplings given by
\eqref{eq:int2} at each input $i$ and $\bsy$. Then, the results in
\ref{tab1} apply for both the inner and outer integrals. If we use an
$n$-point rectangle rule for the inner integral and an $N$-point lattice
cubature rule for the outer integral, and if the functions $f_{i,j}$ are
periodic (thus so are the functions $g_i$), then we will fall under
Scenarios (A5)--(A7) for the inner and outer integrals. Furthermore, if
all functions $f_{i,j}$ are the same, then the final cost is of order
\[
   N\,\log(N) + N\, n\,\log(n)\,.
\]
The cost is independent of $L$, while the error is of order $N^{-\alpha} +
n^{-\alpha}$, with $\alpha$ given by the smoothness of the functions
and the underlying lattice rule, with an implied error constant that may depend
exponentially on $L$.

\subsubsection*{STAGE TWO}

The next two lemmas show that the problem can be simplified further for
this particular model with periodic functions $f_{i,j}$. The proofs can be
found in \cite{HJKLNS21}.

\begin{lemma}[\cite{HJKLNS21}] \label{lem1}
Assume the functions $f_{i,j}$ are periodic. Then the inner integral
\eqref{eq:inner} simplifies to
\begin{align} \label{eq:inner2}
  g_i(\bsy)
  \,=\, g_i\Big(\textstyle\sum_{j=0}^{L-1}  y_j,0,\ldots,0\Big)
  \,=\, g_i\Big(\textstyle\sum_{j=0}^{L-1}  y_j,\bszero\Big),
\end{align}
that is, $g_i(\bsy)$ depends only on the sum of the components of $\bsy$.
\end{lemma}

\begin{lemma}[\cite{HJKLNS21}] \label{lem2}
Assume the functions $f_{i,j}$ are periodic. Then the outer integral
\eqref{eq:outer} simplifies to
\begin{align} \label{eq:outer2}
   \calI \,=\,
  \int_{D^L} \prod_{i=0}^{L-1} g_i\big(y_{i+1} - y_i,\bszero\big)\,\rd\bsy.
\end{align}
\end{lemma}

The last transformation in Lemma~\ref{lem2} leaves the outer integral
\eqref{eq:outer2} in the form of \eqref{eq:int2} and so there is no longer
a need to use a lattice cubature rule. By taking now an $n$-point
rectangle rule for both the inner and outer integrals, Scenario~(A7)
applies in both cases and the cost becomes
\[
   n^2\,\log(n)\,.
\]
The cost is again independent of $L$. The approximation error is then
$\calO( n^{-\alpha})$, where $\alpha$ depends on the smoothness of the
functions, but importantly, the implied constant no longer depends on $L$.

\subsubsection*{STAGE THREE}

An alternative approach based on Fourier series was used in
\cite{HJKLNS21} to simplify the expression even further, as shown in the
following theorem. The proof can be found in \cite{HJKLNS21}.

\begin{theorem}[\cite{HJKLNS21}] \label{thm}
Suppose that the functions $f_{i,j}$ are periodic and have absolutely
convergent Fourier series. Define $\mu_{i+jL} := f_{i,j}$ for
$i,j=0,\ldots,L-1$. Then the integral \eqref{eq:outer}, with inner
integral \eqref{eq:inner}, simplifies to
\begin{align} \label{eq:outer3}
   \calI \,=\,
  \int_{D^{L^2}} \prod_{k=0}^{L^2-1} \mu_k\big(x_{k+1}-x_k\big)\,\rd\bsx,
\end{align}
where now the parametric periodicity is to be taken modulo $L^2$, i.e.,
$x_k \equiv x_{k\bmod L^2}$.
\end{theorem}

Theorem~\ref{thm} shows that, instead of nested integrals, we now have a
single integral \eqref{eq:outer3} with dimensionality $L^2$ of the form
\eqref{eq:int2}, with $L$ replaced by $L^2$. We are again in Scenario~(A7)
with an $n$-point rectangle rule and the cost is only of order
\[
  n\,\log(n)\,,
\]
and the error is $\calO(n^{-\alpha})$. Both the cost and the error bound
are independent of $L$.

\section{Summary} \label{sec:summary}

This proceedings article reviews our recent work published in
\cite{HJKLNS21}. We presented very high-dimensional integrals exhibiting a
special strucure, that arise from models such as the quantum rotor and the
2D compact $U(1)$ LGT. The numerical methods presented here and in
\cite{HJKLNS21} extend the ones presented in
\cite{AGHJLV16,Hartung:2020uuj}. In particular, here and in
\cite{HJKLNS21} we investigated in detail the advantages of applying FFT
and gave sufficient conditions in order to meet the requirements of FFT
application. The latter conditions may apply to other LQCD models than the
ones presented in this work. Executable Julia codes were provided in
\cite{HJKLNS21} together with some numerical results.

The use of lattice cubature rules can be beneficial and even essential
when we have an $L$-fold product of $s$-dimensional integrals, in order to
meet the sufficient conditions to apply FFT. These improvements are not
limited to systems with first order couplings. As shown in \cite{HJKLNS21}
they can be extended to systems with higher order couplings. In the latter
case, the problems can still be truly high dimensional.

The application examples of the quantum rotor and 2D compact $U(1)$ LGT
are encouraging us to look more deeply and to further investigate
potential advantages on truly difficult problems in 3D and 4D compact
$\SU(N)$ LGT. We are currently investigating the latter problems using
recursive numerical integration techniques.

\section*{Acknowledgments}
We gratefully acknowledge financial support from the
Australian Research Council (ARC) under grant DP210100831 and the Research
Foundation Flanders (FWO) under grant G091920N.

\bibliographystyle{plain}

\end{document}